# Angular resolution measurements at SPring-8 of a hard X-ray optic for the New Hard X-ray Mission

D. Spiga[1*], L. Raimondi[1,2], A. Furuzawa[3], S. Basso[1], R. Binda[4], G. Borghi[4], V. Cotroneo[5], G. Grisoni[4], H. Kunieda[3], F. Marioni[4], H. Matsumoto[3], H. Mori[3], T. Miyazawa[3], B. Negri[6], A. Orlandi[4], G. Pareschi[1], B. Salmaso[1], G. Tagliaferri[1], K. Uesugi[7], G. Valsecchi[4], D. Vernani[4]

[1]INAF/Brera Astronomical Observatory, Via E. Bianchi 46, 23807 Merate (LC), Italy
[2]Universita' degli Studi dell'Insubria, Via Valleggio 11, 21100 Como (CO), Italy
[3]University of Nagoya, Furo-cho, Chikusa, Nagoya 464-8602, Japan
[4]Media-Lario Technologies, Località Pascolo, 23842 Bosisio Parini (LC), Italy
[5]Center for Astrophysics, 60 Garden Street, Cambridge (MA) 02138-1516, United States
[6]Agenzia Spaziale Italiana, Viale Liegi 26, 00198 Roma, Italy
[7]Japan Synchrotron Radiation Research Institute, Sayo-cho, Sayo-gun, Hyogo 679-5198, Japan

## ABSTRACT

The realization of X-ray telescopes with imaging capabilities in the hard (> 10 keV) X-ray band requires the adoption of optics with shallow (< 0.25 deg) grazing angles to enhance the reflectivity of reflective coatings. On the other hand, to obtain large collecting area, large mirror diameters (< 350 mm) are necessary. This implies that mirrors with focal lengths ≥10 m shall be produced and tested. Full-illumination tests of such mirrors are usually performed with on-ground X-ray facilities, aimed at measuring their effective area and the angular resolution; however, they in general suffer from effects of the finite distance of the X-ray source, e.g. a loss of effective area for double reflection. These effects increase with the focal length of the mirror under test; hence a "partial" full-illumination measurement might not be fully representative of the in-flight performances. Indeed, a pencil beam test can be adopted to overcome this shortcoming, because a sector at a time is exposed to the X-ray flux, and the compensation of the beam divergence is achieved by tilting the optic. In this work we present the result of a hard X-ray test campaign performed at the BL20B2 beamline of the SPring-8 synchrotron radiation facility, aimed at characterizing the Point Spread Function (PSF) of a multilayer-coated Wolter-I mirror shell manufactured by Nickel electroforming. The mirror shell is a demonstrator for the NHXM hard X-ray imaging telescope (0.3 – 80 keV), with a predicted HEW (Half Energy Width) close to 20 arcsec. We show some reconstructed PSFs at monochromatic X-ray energies of 15 to 63 keV, and compare them with the PSFs computed from post-campaign metrology data, self-consistently treating profile and roughness data by means of a method based on the Fresnel diffraction theory. The modeling matches the measured PSFs accurately.

**Keywords:** Hard X-ray optics, NHXM, SPring-8, optics calibration

## 1. INTRODUCTION

Future X-ray telescopes like the New Hard X-ray Mission (NHXM[1]), NuSTAR[2], and ASTRO-H[3] will enable the observation of the Universe in the hard (> 10 keV) X-ray band. The imaging and polarimetric NHXM telescope, in particular, is aimed at extending to the hard X-ray band the high angular resolution achieved by the Newton-XMM[4] (15 arcsec HEW - *Half-Energy-Width* - at 1 keV) and SWIFT/XRT[5] (17 arcsec HEW at 1 keV). Their imaging capabilities are made possible by their performing optical systems, consisting of grazing incidence mirrors obtained by replication of Wolter-I[6] superpolished mandrels by Nickel electroforming[7]. This technique has been proven to yield mirrors with high accuracy, not only in terms of mirror figure, but also in surface microroughness, an important aspect in order to limit the imaging degradation caused by the X-ray scattering[8] (XRS), which in general exhibits a strong dependence on the X-ray energy. At this regard, the NHXM requirement[1] is to achieve a HEW better than 20 arcsec

---

[*] contact author: daniele.spiga@brera.inaf.it, tel. +39-039-5971027, fax +39-039-5971001

below 30 keV, with an effective area extended up to 80 keV. The current optical design[9] of the NHXM optical system is based on 4 mirror modules with a 10 m focal length, each of them including 70 mirror shells with diameters up to of 350 mm, coated with Pt/C graded multilayers to endow them with X-ray reflectivity up to 80 keV. An extended design would consist of adding 20 innermost shells coated with Ni/C multilayers[10] to bring the maximum energy of sensitivity to 120 keV. A noticeable effort[11] has been made at *INAF/Brera Astronomical Observatory* (OAB) and *Media-Lario Technologies* (MLT) to upgrade the Nickel electroforming technique in order to maintain a good angular resolution in hard X-rays with a mirror thickness-to-radius ratio smaller than XMM's by a two-fold factor, and in spite of the stress induced by the multilayer coating. Prototypes of NHXM mirror modules with a few mirror shells were manufactured[9], aiming at demonstrating the feasibility of mirrors with such angular resolutions, e.g., adopting an electroformed Nickel-Cobalt alloy, stiffer than pure Nickel[11], and reducing the thickness of the gold layer used for the mirror release to improve the roughness[12]. The direct performance verification was done by measuring the X-ray PSF (*Point Spread Function*) up to 50 keV in full-illumination setup at PANTER (MPE, Germany)[13], a reference X-ray facility for calibration of X-ray telescopes like ROSAT, Beppo-SAX, JET–X, SOHO/CDS, ABRIXAS, Chandra LETG, XMM, SWIFT/XRT, Suzaku.

However, testing double reflection mirrors with long focal lengths, in full illumination setup, requires extremely collimated and wide X-ray beams. At PANTER, this is obtained by locating the X-ray source at a 120 m distance from the module under test. Even so, the finite distance effects are clearly felt. The effective area loss represents a major problem: rays reflected at points close to the parabola front-end *miss the second reflection and are not focused at the correct distance* (Fig. 1, top). For mirrors with large f-numbers, the ratio of double-reflected rays[14] only depends on the focal length, $f$, and the distance of the source, $D$:

$$V = \frac{D - 4f}{D + 4f} . \tag{1}$$

For short focal lengths like SWIFT/XRT (3.5 m), this is not a problem because > 80% of the mirror surface is seen in double reflection. For a NHXM mirror tested at PANTER this ratio descends to 50%. Therefore, a HEW measurement is only representative of half length of the primary segment. If the other half had a worse shape/roughness, it would not be seen in full illumination, but would significantly degrade the PSF once illuminated by an astronomical X-ray source.

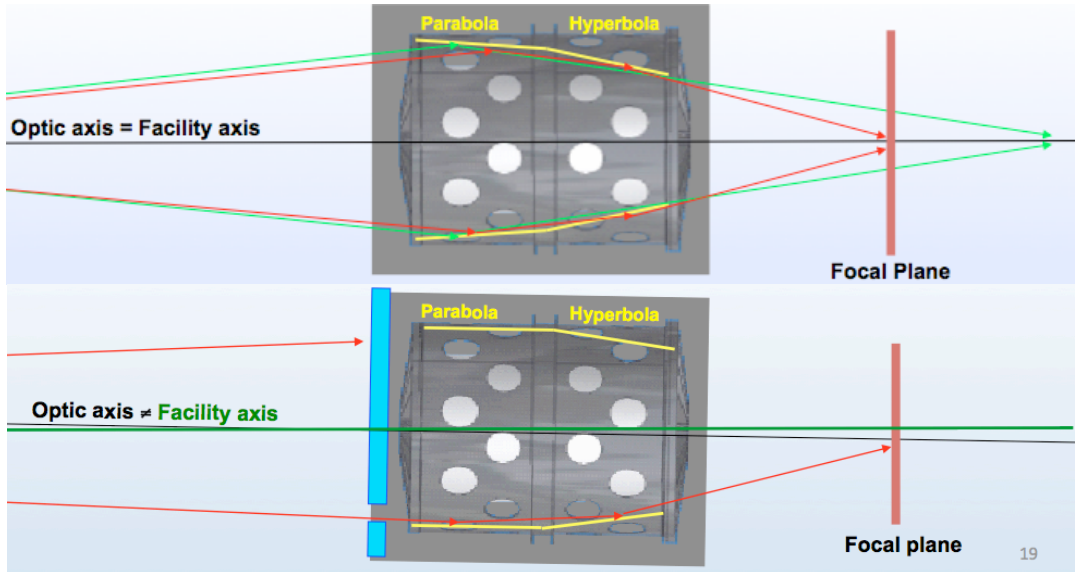

Fig. 1: (top) *Full illumination setup*: rays hitting the primary front-end miss the second reflection; hence only a part (Eq. 1) of the mirror surface is seen in double reflection. (bottom) *Pencil beam setup*: a sector at a time is illuminated and tested. The lateral tilt corrects the effect of the source at finite distance.

Even if PANTER tests in full illumination are very useful to directly measure the PSF of 50% of a NHXM mirror shell, it is interesting to perform an independent PSF measurement aimed at characterizing the entire length of the primary mirror. This can be done adopting a *pencil-beam setup* as in Fig. 1 (bottom): by means of an opaque mask, only a small sector at a time is illuminated. The mirror is then tilted laterally to compensate the divergence, nulling thereby the stray light from the parabola. The finite distance of the source still results in a slight blurring of the image (~ 1 arcsec for NHXM at PANTER) and in a focal length displacement[6] (+ 90 cm for the same case), but the entire length of the parabolic segment is seen in double reflection. By spinning the optic about its axis, all the sectors can be characterized and the PSF be reconstructed by superposing the individual exposures. This pencil beam technique has already been adopted for previous X-ray missions, e.g. ASCA[15]. Another technique that could be adopted foresees the diffraction of X-rays off a Silicon crystal in asymmetric configuration[16], as it was done to calibrate the SODART telescope at the Daresbury synchrotron[17].

Even if an adoption of the pencil beam at PANTER[14] is foreseen to calibrate the NHXM optics after completion of an appropriate manipulator[9], we envisaged conducting parallel measurements with synchrotron light at the beamline BL20B2 of the *SPring-8 radiation facility* (JASRI, Hyogo prefecture, Japan). The reason is that SPring-8 is a powerful source with a copious emission flux extended to more than 110 keV, thereby minimizing the exposure time per tested sector. Although BL20B2 was mainly designed for medical imaging, it has been equipped with a manipulating drum and specific imaging detectors to perform the calibration in pencil beam of the X-ray telescope ASTRO-H[18],[19],[20], therefore it is suitable to mount also the NHXM module prototype via a flange serving as a mechanical interface to fit the different module diameter. Also the X-ray telescopes onboard InFOCµS and SUMIT have been calibrated at this beamline[18].

In this paper we report the results obtained in the characterization campaign at SPring-8 on the Technical Demonstrator Model No. 2 (TDM2) for the NXHM project. Even though the module included three mirror shells with diameters 185, 297, and 350 mm, representative of the smallest, the average, and the largest mirrors of the NHXM module, the tests were conducted only on the mirror shell with 297 mm diameter (heretofore, MS297), which exhibited the best performances at PANTER tests. The features of this shell are listed in Tab. 1: details on the MS297 mirror development[11] and the results of the PANTER tests can be retrieved from another paper of this volume[9]. The experimental setup is briefly described in Sect. 2. In Sect. 3 we show the reconstructed PSF of the MS297 at monochromatic energies in the range 15 – 63 keV and the HEW values we derived. Sect. 4 displays the profile and roughness measurements achieved at MLT and INAF/OAB after the characterization at SPring-8. Metrology data are self-consistently used in Sect. 5 to simulate the PSFs at the X-ray energies probed at SPring-8: the simulated PSFs are in very good agreement with the results of the pencil beam characterizations. Results are summarized in Sect. 6.

| | |
|---|---|
| *Nominal focal length* | 10.037 m |
| *Mirror diam. at the intersection plane* | 297 mm |
| *Mirror length (2L)* | 600 mm |
| *Mirror wall thickness* | 0.26 mm |
| *On-axis incidence angle* | 0.21 deg |
| *Number of sectors* | 24 |
| *Multilayer coating* | 95 (20+75) bilayer graded W/Si |
| *HEW measured in UV light (218 nm)* | 21 arcsec (aperture diffr. subtr.) |
| *HEW meas. at PANTER (1 keV)* | 18 arcsec |
| *HEW meas. at PANTER (30 keV)* | 24 arcsec |

Tab. 1: parameters of the Wolter-I mirror shell[9] tested at Spring-8.

## 2. EXPERIMENTAL SETUP AT SPRING-8 BL20B2

SPring-8 is the world's largest 3$^{rd}$ generation synchrotron light source. The beamline BL20B2 (Fig. 2) is 215 m long from the bending magnet to the end of the experimental hutch, leaving 201.203 m from the X-ray source (150 µm hor. size x 10 µm vert. size) to the principal plane of the optic. The remaining 14.8 m are sufficient to accommodate the 10.523 m distance to the detector (the distance is larger than the 10 m focal length because of the finite distance of the source). A Silicon double crystal is used to monochromate the radiation at 5 to 113 keV. We have used the (311) reflection, enabling the selection of X-ray energy in the 8.4 – 72.5 keV band, with a ΔE/E resolution of 10$^{-4}$. The beam enters the experimental hutch No. 2 through a Kapton window and impinges the mirror under test. The beam is very intense ($1.5 \times 10^7$ count/sec/mA/mm$^2$ at 30 keV, at the hutch No. 3) and collimated: less than 1 arcsec per mm of entrance slit width, in the horizontal direction. The vertical divergence is negligible for our scopes. The maximum available beam size is 300 mm in horizontal and 22 mm in vertical[18].

The TDM2 optic was mounted at BL20B2 in front of the Kapton window (Fig. 3). The manipulating drum was suited to perform an accurate alignment of each mirror sector with respect to the X-ray beam and to spin the optic about its axis, exposing different sectors to the X-ray flux. The annular (projected) width of the mirror parabolic segment (1 mm) determined the beam divergence, which resulted in a collimation within 1 arcsec over the sector surface. The vertical size of the used slit was 20 mm, which entirely fitted the height of a sector (35 mm) between two consecutive spokes of the spider, without obstructions. The exact horizontal size (5 mm) was not relevant, provided that it was sufficiently wide to include the annular width of the primary segment (Fig. 4).

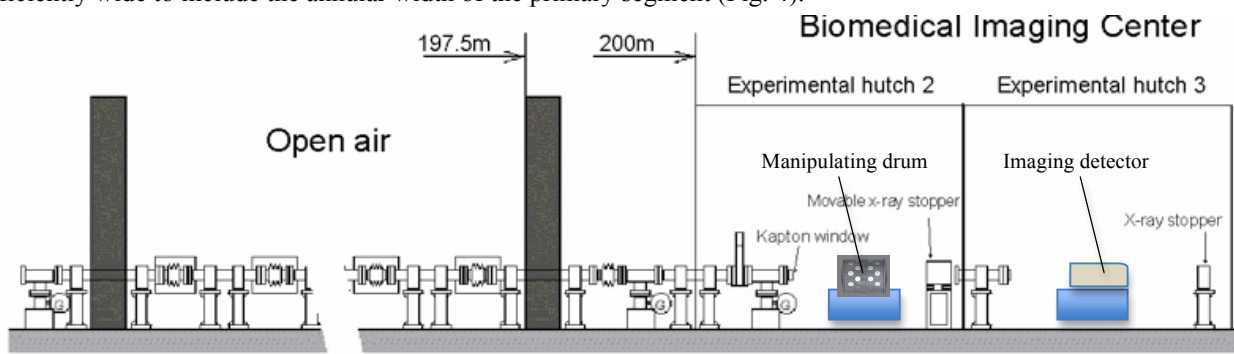

Fig. 2: the setup of the beamline at the beam exit (after *http://www.spring8.or.jp/*). The mirror module under test was located in front of the Kapton window in the hutch No. 2. The detector was set in the experimental hutch No.3, 10.52 m away from the optic intersection plane.

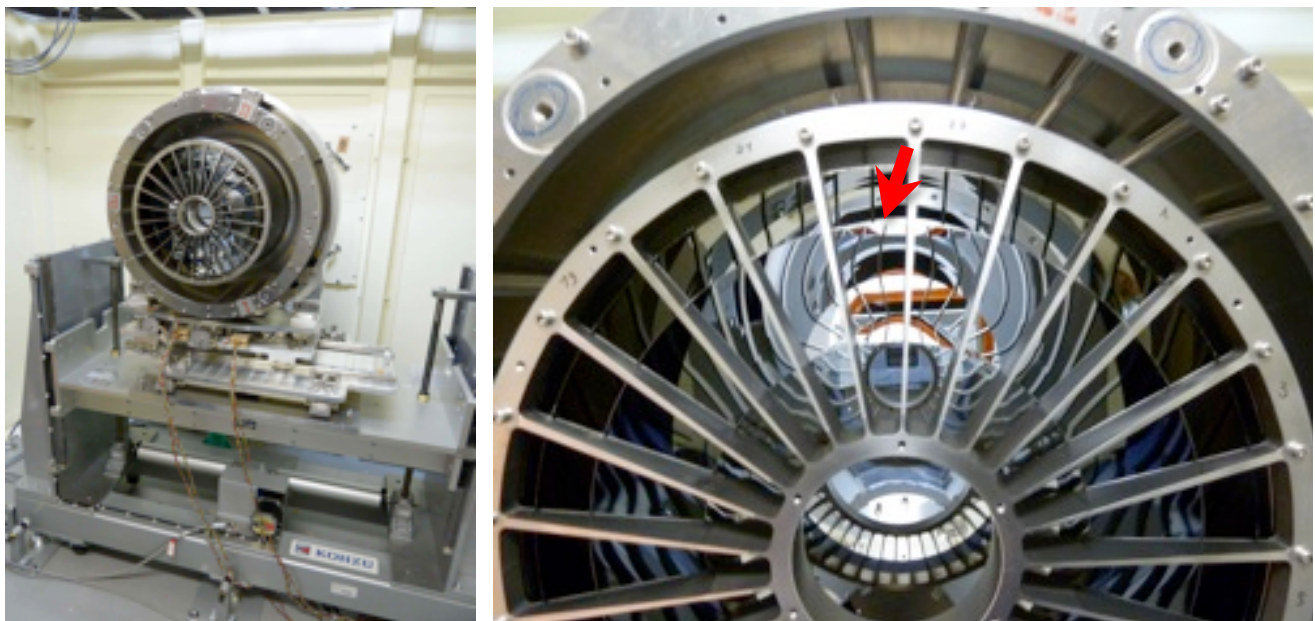

Fig. 3: (left) the optical module mounted in the manipulating drum at BL20B2. (right) close view on the TDM2. The arrow points to (the outer surface of) the tested mirror shell, the MS297.

After selection of a monochromatic X-ray energy in the 15 to 63 keV energy range, a single sector of the MS297 was located in front of the beam (Fig. 4). The uniformity of the beam intensity was 3% in the horizontal direction; therefore an exact positioning of the mirror sector was not crucial. The reflected beam was then recorded using a 45 mm × 30 mm Scintillator+CCD imaging detector with 11.3 µm-sized pixels. Aiming at speeding the readout up, every second pixel was read, resulting in an effective pixel size of 22.6 μm, which at a 10.5 m distance is equivalent to a 0.5 arcsec resolution of the detector. The alignment of every sector was achieved by varying the tilt and rocking angles of the optic, until the stray light from the parabolic segment just disappeared completely. Actually, because the spin axis of the manipulator did not coincide with the optic axis, the optic did not maintain the correct orientation of rays when the sector was changed, hence we had to check and refined the alignment of each sector before recording the focus image. After the alignment, the focal spot of a single sector looked like in Fig. 5.

The first measurement at 30 keV was also addressed to the search for the best focus, i.e., the detector distance from the mirror at which the intensity distribution for a single exposure exhibits the most symmetric distribution on the left and the right side (the method suited in full illumination, i.e., the HEW minimization, would not have worked because it is mostly sensitive to the simultaneous convergence of the focused beam from *all* the sectors). After subtracting the effect of the distance of the X-ray source at BL20B2, the measured focal length of the mirror is 10.05 m, in good accord with measurements in UV light and X-rays in full illumination, and consistently with the nominal focal length (Tab. 1) within the measurement uncertainties. The copious photon flux allowed us integrating for a very short time with excellent statistics, therefore *all* sectors have been aligned and exposed at several X-ray energies: 15, 20, 30, 35, 40, 45, 50, 55, 60, 63 keV. The exposure time varied with the reflectivity, from 70 msec (20 keV) to 3 sec (60 keV), to avoid pixel saturation. The detector noise was measured in separate "dark" exposures.

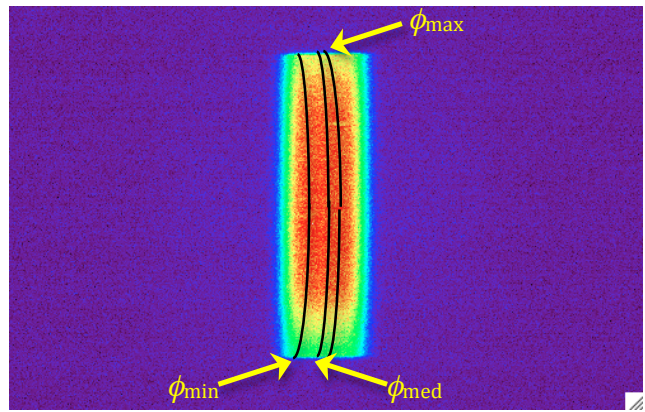

Fig. 4: the direct beam emerging from the 2 cm-high slit. We have superposed to the image the *approximate* position of the maximum, the median, and the minimum diameter of the Wolter-I MS297, once aligned. The collecting area is the zone between the maximum and the median diameter.

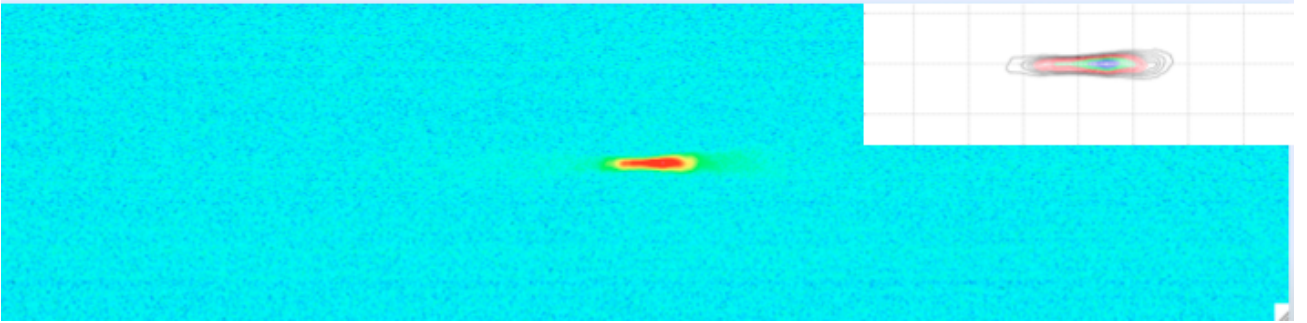

Fig. 5: the focused image of a single exposure after sector alignment. The image width is 4 cm (the high and the low part of the images are trimmed out). We also show a contour plot of the image. The intensity distribution exhibits a sharp peak.

## 3. RECONSTRUCTED HARD X-RAY PSF AND HEW

From the exposures of the individual sectors at the best focus we have reconstructed the PSF at each the X-ray energy. To do this, we have preliminarily subtracted from every image the dark exposure at the same energy. However, because the noise level might be fluctuating in time, we have selected a 200 pixel-high, 4 cm wide box in the central region of every image. We have then subtracted another noise map, obtained as the average of two horizontal slices of the same size taken from the top and bottom of the image, which should represent the noise of the CCD at the measurement time.

Finally, we selected a 100 pixel-high region, centered on the focal spot without changing the width, and we performed a final subtraction of the noise using the 50 pixel-high trimmed regions, joined together. In this way, we have removed any fluctuation of the average noise level that can affect the estimation of the PSF.

After noise removal, we have rotated the resulting image strips and superposed them. Since the axes mismatch may have introduced a lateral displacement of the sectors, which we could not measure, the center of rotation was not kept at fixed coordinates, but fluctuated by a few pixels. Therefore, we set the rotation center at the maximum count position in every exposure. This choice suppressed the impact of roundness errors on the PSF, but preserves the angular spread due to longitudinal profiles and scattering, which are expected to be the dominating terms. The resulting PSFs do not sample the entire mirror surface, because the slit height was smaller than that of the sector, but the measurement is representative of a large fraction (63%) of the mirror effective area. The missing regions correspond to gaps in the azimuthal angle, but are representative of the entire mirror length. At PANTER, in contrast, the azimuthal angle is mapped completely at the expense of the primary mirror length, which is not focused by 50%. In short, the two characterizations are complementary. The reconstructed focal spots at some selected energies are displayed in Fig. 6.

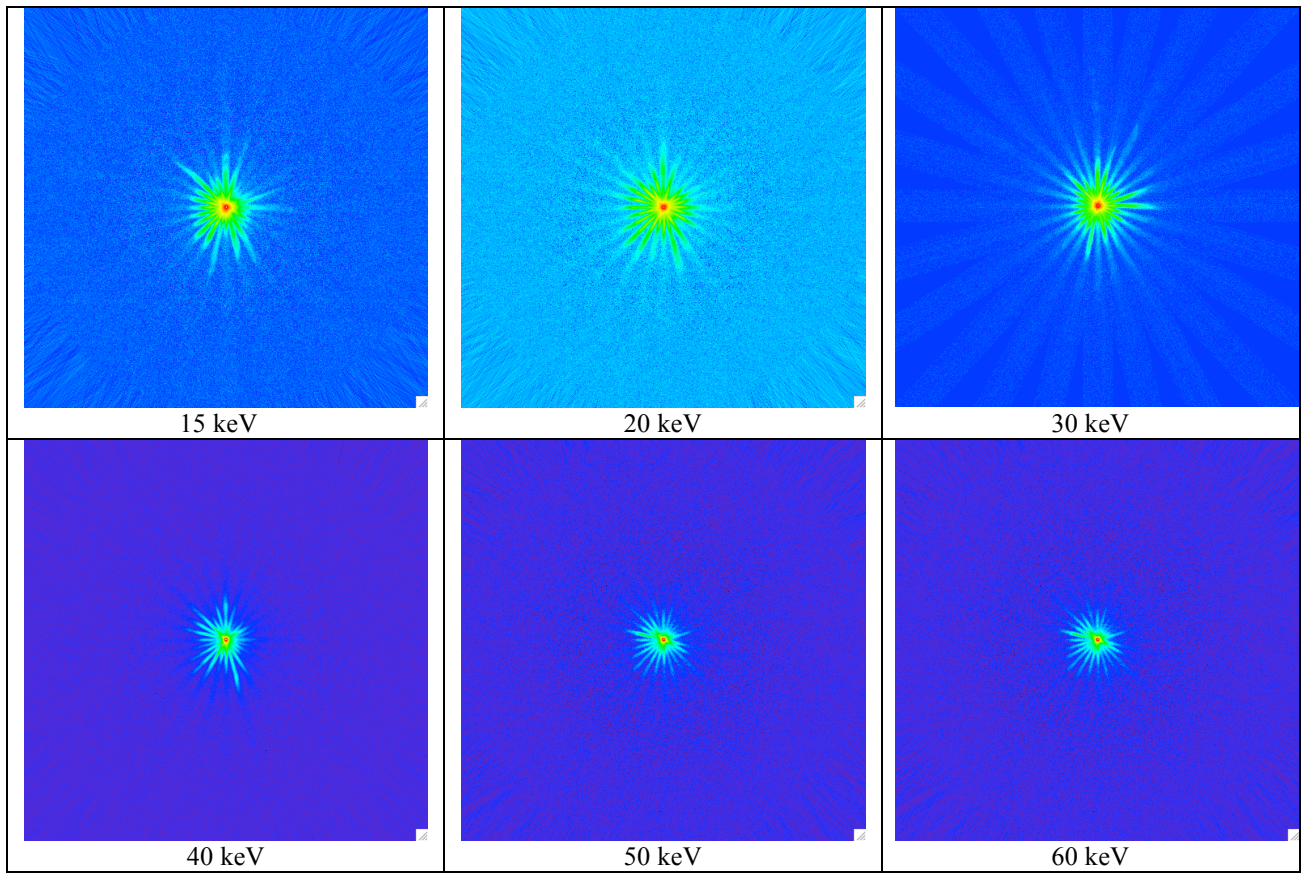

Fig. 6: the reconstructed focal spot of the MS297, at increasing energies. All images appear to be consistent with each other. The focal spot *seems* to shrink because of the decreasing reflectivity that causes the focused beam to be less prominent in the background (even though the background is nearly reduced to zero *on average*).

The mirror shell PSFs have been derived from reconstructed images by annular integration, and the HEW values have been computed. The HEW trend, clearly increasing with the X-ray energy especially beyond 40 keV, is shown in Fig. 7. A comparison with the corresponding HEW values measured at PANTER, reported in another paper of these proceedings[9], shows that the results of PANTER and SPring-8 are consistent since the trends are quite well superposed in the 15-40 keV common energy range. At 30 keV, however, the HEW measured at SPring-8 is sensitively larger (28 arcsec Vs. 24 arcsec measured at PANTER). The difference can be ascribed to the different tested region of the mirror, in particular the better value found at PANTER might denote a better roughness of the parabolic segment near the intersection plane. In the next section we will see that the metrology results, i.e., the measured profiles and roughness of the MS297, explain the measured PSFs accurately.

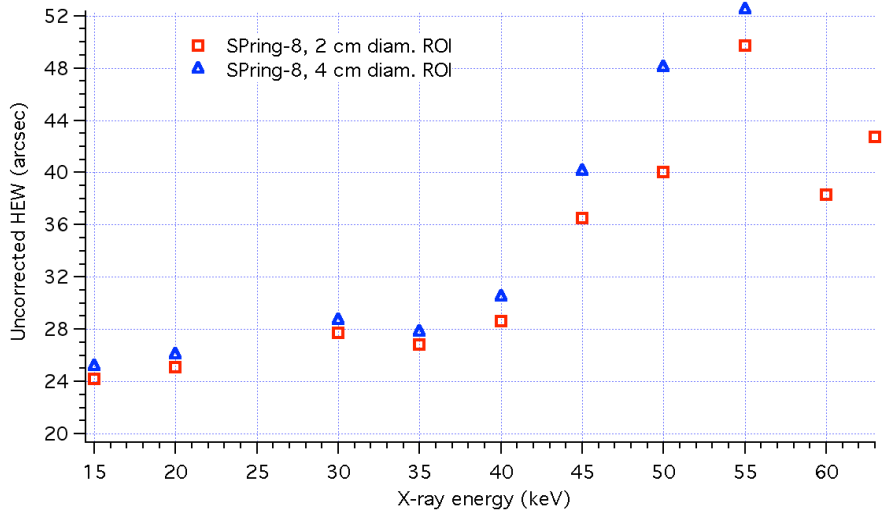

Fig. 7: the HEW values of the MS297, measured at SPring-8 BL20B2. The slightly higher values for a wider region of interest (ROI) denote some scattering of X-rays also at the detector's edge. The results comply the measured HEW measured at PANTER at 15 to 40 keV, reported in another paper of this volume[9].

## 4. MIRROR METROLOGY: EXPLANATION OF THE MEASURED PERFORMANCES

After the measurements performed at Spring-8 BL20B2, the MS297 has been characterized in profile and roughness in order to explain the focusing performances observed in X-rays, especially at high energies. Usually, figure errors determine the PSF at low energies (~1 keV) and can be treated with geometrical optics methods (i.e., ray-tracing). On the other side, roughness covers the high frequency range of mirror imperfections, and need to be treated with the X-ray scattering theory, which contributes to degrade the PSF to an extent increasing with the X-ray energy[8]. This simplified view supposes an abrupt boundary between the two regimes, which is neither abrupt nor easy to locate. Nevertheless, we will later see that this approach is correct in the case of this mirror: longitudinal profiles can then be used to derive a figure error term of the HEW, whilst roughness measurements, once processed in terms of *Power Spectral Density* (PSD), can be used to derive the HEW scattering term, as a function of the X-ray energy[21].

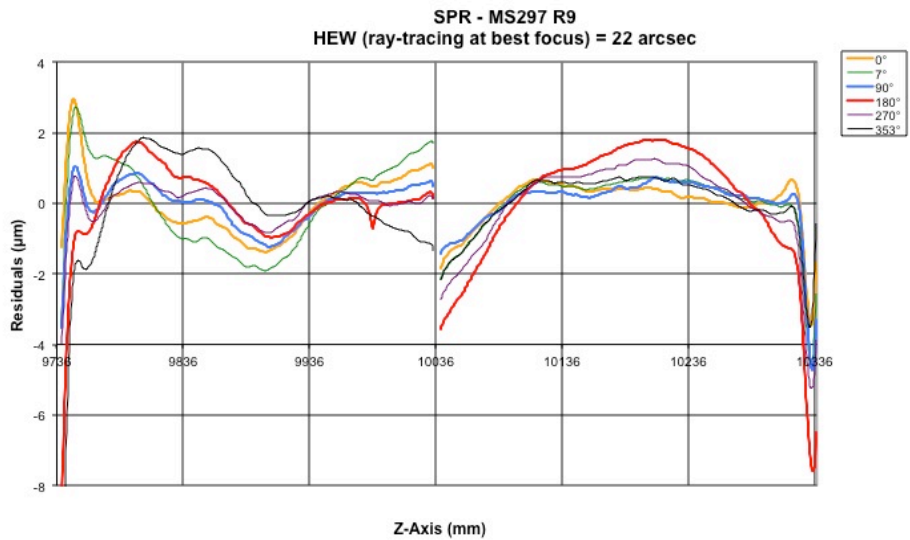

Fig. 8: some longitudinal profiles of the MS297, as measured with the Shell Profilometer/Rotondimeter (SPR[22]) operated at MLT (0.4 mm lateral resolution). The graph shows the residuals of the 6 measured profiles with respect to the nominal Wolter-I. The hyperbola is on the left side, the parabola on the right side.

Six longitudinal profiles of the MS297 (Fig. 8) have been measured using the SPR (Shell Profilometer/Rotondimeter) specifically developed and operated at MLT for the NHXM project[22]. The SPR enables high-accuracy profile measurements of mirror shells without the need of removing them from the integration case. Actually, for this specific case the innermost shell (MS185) had to be removed, therefore the profile measurement of the MS297 was taken without the upper spider. This might potentially have slightly changed the shape of the mirror, even if the integration process is conceived and proven to not change the mirror profile, hence the spider removal could have changed the HEW only by a few arcsec. In fact, the HEW resulting from ray-tracing the profiles is 22 arcsec, Vs. 18 arcsec measured at PANTER at 1 keV[9]. This overestimate may also be caused by the incorrect application of geometrical optics to the highest spatial frequencies present in the measurement (i.e., close to (0.4 mm)$^{-1}$).

After the profile characterization, the MS297 was removed and cut into pieces to perform roughness measurements at INAF/OAB. High-frequency roughness measurements were performed using a stand-alone *Veeco Explorer®* Atomic Force Microscope (AFM) at 3 different magnifications (Fig. 9) down to a maximum lateral resolution of 2.5 nm. The most noticeable feature, especially in the 10 µm scan, is the presence of small dots spread over the surface. These defects were already detected on the mandrel as small pores, filled by the gold layer and replicated by the mirror shell[12]. AFM measurements cover the spectral band 50 µm – 5 nm of spatial wavelengths. Measurements via PSI (Phase Shift Interferometry) to characterize the roughness at larger spatial wavelengths (1 mm - 50 µm) could not be performed, due to the sample curvature. Based on previous experiences, we have assumed in this region the same roughness of the mandrel, which was characterized completely prior to mirror replication.

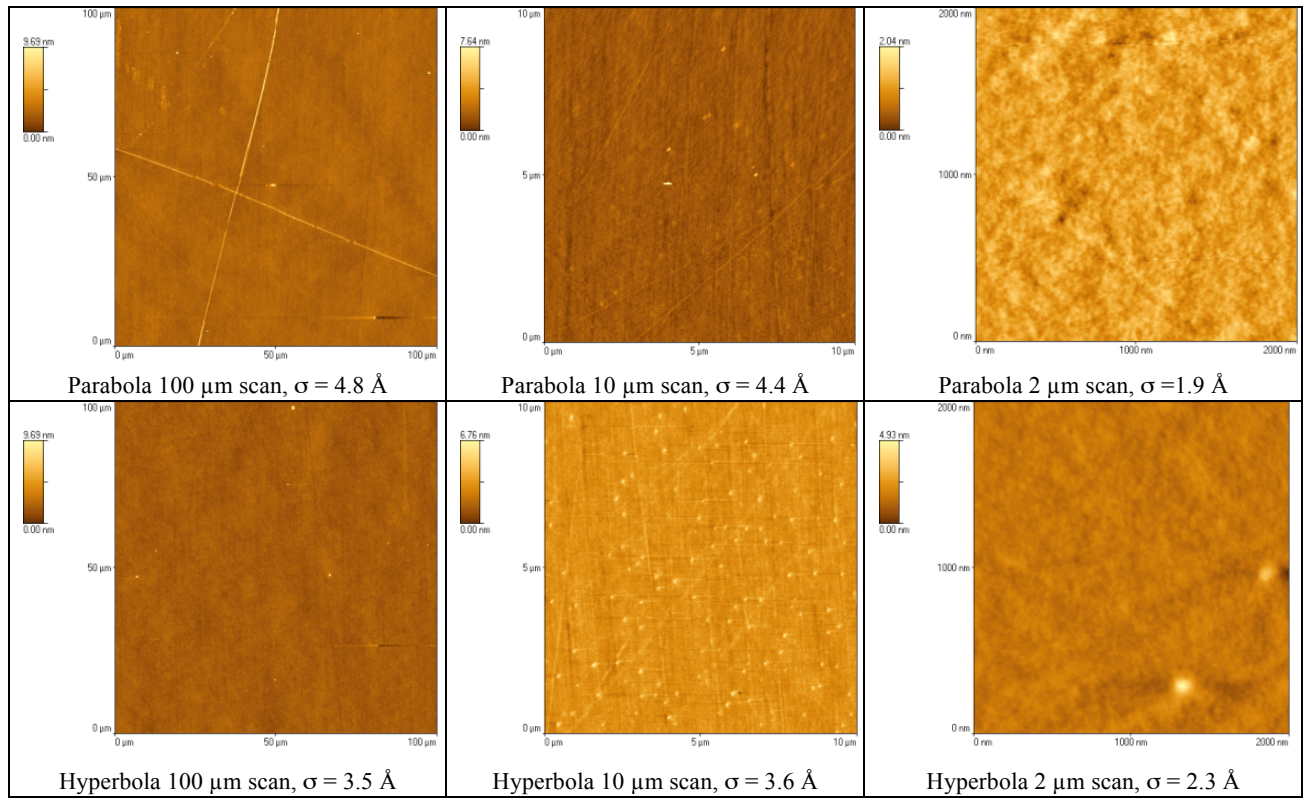

Fig. 9: AFM images of the MS297 surface. The dots visible on the surface are more evident on the hyperbolic segment, but they are also present on the parabola, even though they are less visible because they are confused among other defects.

The complete roughness characterization of the MS297 is reported in Fig. 10. We have computed the average PSD from the AFM maps, and compared them to that of the mandrel post 5$^{th}$ replica. The PSD of the hyperbola is very well superposed to that of the mandrel excepting for wavelengths below 0.1 µm, where the shell is slightly rougher. This suggests that the multilayer deposition did not increase the surface roughness significantly. The parabola surface, in contrast, is sensitively rougher in all the AFM range. As we mentioned, the roughness of the mirror could not be measured with PSI techniques: however, we can reasonably assume that the PSD of the shell in the sub-millimeter range is the same as that of the mandrel. In fact, *the PSD can be indirectly derived from the HEW values measured at*

*SPring-8,* using the inverse analytical formalism[21] relating the XRS term of the HEW, as a function of the X-ray energy, to the PSD of the mirror. The PSD computed in this way is in the range 1 mm – 70 μm of spatial wavelengths and matches well the PSD of the mandrel, on condition that 18 arcsec of figure error HEW (i.e., the measured HEW value at 1 keV) are *linearly* subtracted from the measured HEW trend. We thereby obtain that the HEW increase measured at SPring-8 is justified assuming that the mirror surface PSD has replicated the topography of the mandrel in the 1 mm – 70 μm spatial wavelength range.

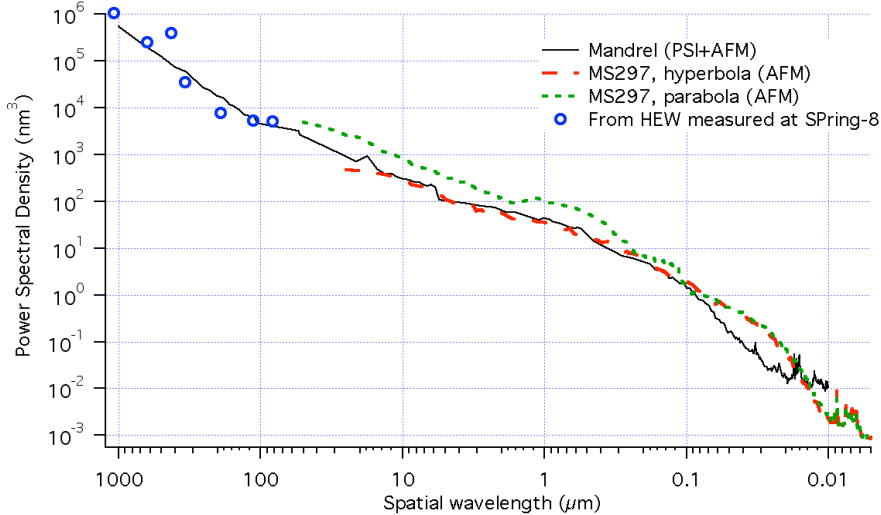

Fig. 10: Power Spectral Density characterization of the MS297. The parabolic segment has nearly the same finishing level as the mandrel. The hyperbolic segment is rougher. The PSD in the PSI spectral range is obtained from the HEW values measured at SPring-8, assuming a 18 arcsec HEW to be subtracted linearly from the measured trend.

Using the measured profiles (covering the perturbation spectrum down to 0.8 mm) and the measured roughness PSD (at 1 mm to 5 nm spatial wavelengths), we can now simulate the Point Spread Function (PSF) at any X-ray energy. To this end, we make use of a method based on the Huygens-Fresnel principle, previously developed for single-reflection mirrors[23]: the extension to the case of a Wolter-I mirror is described in another paper of this volume[24]. This formalism enables the self-consistent computation of the PSF of a Wolter-I mirror, given the profile and PSD characterization, without the need of setting any boundary between geometric errors and surface roughness. In Fig. 11 we compare the HEW values as computed from the simulated PSFs using the Fresnel diffraction theory with the measured HEW at SPring-8. We also show the HEW trend computed analytically from the PSD[21], after adding *linearly* 18 arcsec of figure error, as we assumed in the inverse computation of Fig. 10. Data match to within a few arcseconds.

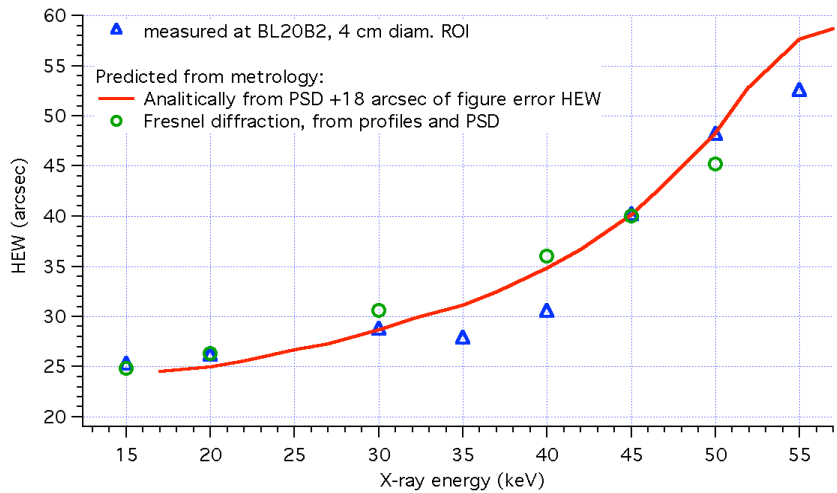

Fig. 11: MS297 HEW trend as measured at SPring-8, compared with the results of the analytical computation (line) and the HEW consistently computed from the Fresnel diffraction theory (circles: see also the PSFs in Figs. 12 through 14).

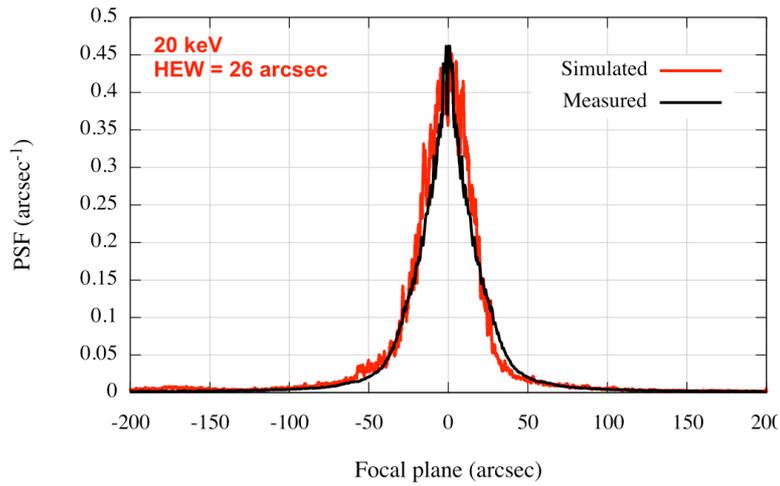

Fig. 12: simulated PSF of the MS297 from measured profiles and roughness, at 20 keV, using the Fresnel diffraction approach[24]. The experimental PSF is reproduced accurately.

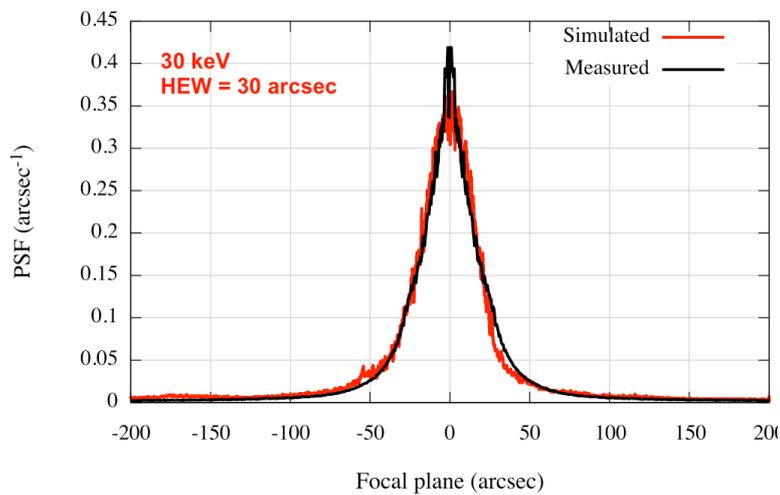

Fig. 13: simulated PSF of the MS297 from measured profiles and roughness, at 30 keV, using the Fresnel diffraction approach[24]. The experimental PSF is reproduced accurately. Notice the broadening and the reduction of the peak value with respect to Fig. 12.

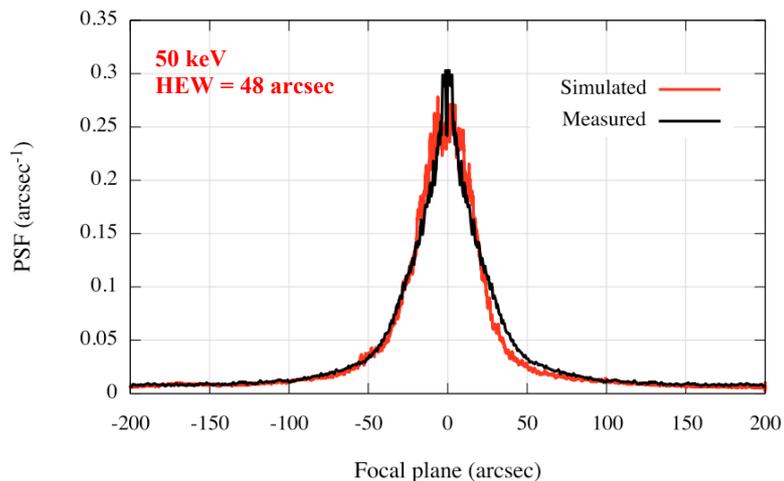

Fig. 14: simulated PSF from measured profiles and roughness, at 50 keV, using the Fresnel diffraction approach[24]. The experimental PSF is reproduced accurately. Notice the broadening and the reduction of the peak value with respect to Figs. 12 and 13.

Finally, in Figs. 12, 13, and 14, we display the simulated PSFs of the MS297 computed from measured profiles and PSD at 20 keV, 30 keV, and 50 keV respectively. The PSFs were obtained by superposing a rough profile, simulated from the PSD (Fig. 10), to each one of the SPR measurements (Fig. 8), and applying to the resulting profile the formulae derived from the Fresnel diffraction theory[24]. The 6 PSFs were then averaged together and normalized over a 4 cm wide region, the same lateral size of the detector used to reconstruct the PSFs (Fig. 6). Exactly the same procedure was applied to the same metrology dataset, varying only the X-ray energy. *The simulations accurately match the measured PSFs computed from the SPring-8 measurements*. Consequently, also the HEW values computed from the simulated PSFs are in good accord with the measured values (Fig. 11) to within a couple of arcseconds. The remaining discrepancies might be caused by the small number of profiles measured with the SPR, which might not be fully representative of the figure of all the tested sectors at SPring-8.

## 5. CONCLUSIONS

The angular resolution measurements of a hard X-ray Wolter-I optic prototype for NHXM performed at Spring-8 have made possible an extension of the characterizations, usually performed at PANTER, to higher energies, not far from the upper limit of the energy range of NHXM (80 keV, in the baseline design). The measurement campaign also allowed us to overcome the problem of the effective area loss due to the finite distance of the source. The measured HEW of the MS297 in pencil beam setup, after a reconstruction of the PSF, is in substantial agreement with the measurements achieved at PANTER[9] at lower energies. In particular, the HEW of the MS297 at 30 keV is close to, although not compliant yet with, the NHXM target of 20 arcsec. The measured HEW increases with the X-ray energy in good agreement with the metrology of profiles and surface roughness, under the condition that the HEWs related to the figure errors and scattering can be added linearly, not in quadrature as usually supposed. Finally, the PSF in hard X-rays *can be correctly predicted by analyzing in a self-consistent way the profile and the roughness, using the Fresnel diffraction approach*. They also match the HEW computed analytically, confirming that, in this case, the XRS impact on the HEW can also be treated separately from the figure error. Therefore, measurements performed at SPring-8 provided an independent confirmation not only of the metrology characterization, but also of the method adopted to derive the PSF from metrology data.

## ACKNOWLEDGMENTS

This work has been financed by the Italian Space Agency (contract I/069/09/0). The synchrotron radiation experiments were performed at the BL20B2 of Spring-8 with the approval of the Japan Synchrotron Radiation Research Institute (JASRI) (Proposal No. 2010B1551).